\documentclass{aa}  

\usepackage{graphicx}
\usepackage{color}
\usepackage{longtable}
\usepackage{supertabular}
\usepackage{natbib}

\definecolor{linkcolor}{rgb}{0.2, 0.55, 0.9}
\usepackage[breaklinks, colorlinks, citecolor=linkcolor]{hyperref}

\usepackage[varg]{txfonts}

\usepackage{listings}
\definecolor{dkgreen}{rgb}{0,0.6,0}
\definecolor{gray}{rgb}{0.5,0.5,0.5}
\definecolor{mauve}{rgb}{0.58,0,0.82}

\usepackage{txfonts}
\begin{document} 
\definecolor{red}{rgb}{1.,0.,0.}
\definecolor{darkgreen}{rgb}{0,0.7,0}

\newcommand{\ms}{M_{\odot}}
\newcommand{\br}{\beta(r)}
\newcommand{\bpr}{\sigma_{\rm r}/\sigma_{\theta}}
\newcommand{\rvir}{r_{200}}
\newcommand{\rtwo}{r_{-2}}
\newcommand{\cvir}{c_{200}}
\newcommand{\vvir}{v_{200}}
\newcommand{\mvir}{M_{200}}
\newcommand{\rnu}{r_{\nu}}
\newcommand{\msun}{M_{\odot}}
\newcommand{\rb}{r_{\beta}}
\newcommand{\rs}{r_{\rm s}}
\newcommand{\ks}{\mathrm{km \, s}^{-1}}
\newcommand{\slos}{\sigma_v}
\newcommand{\slosg}{\sigma_g}
\newcommand{\vrf}{v_{{\rm rf}}}
\newcommand{\ab}[1]{\textcolor{magenta}{\bf ABi: #1}}
\newcommand{\ma}[1]{\textcolor{red}{\bf MAb: #1}}
\newcommand{\jb}[1]{\textcolor{blue}{\bf JBe: #1}}
\newcommand{\new}[1]{\textcolor{magenta}{\bf #1}}

   \title{\texttt{DS+}: a method for the identification of cluster substructures}

   \author{Jos\'e A. Benavides \inst{1,2}\thanks{E-mail: jose.benavides@unc.edu.ar}
          \and          
          Andrea Biviano\inst{3,4}
          \and
          Mario G. Abadi \inst{1,2}        
          }

   \institute{
   Instituto de Astronom\'ia Te\'orica y Experimental, CONICET-UNC, Laprida 854, X5000BGR Córdoba, Argentina
   \and
   Observatorio Astron\'omico de C\'ordoba, Universidad Nacional de C\'ordoba, Laprida 854, X5000BGR C\'ordoba, Argentina
    \and   
      INAF-Osservatorio Astronomico di Trieste, via G. B. Tiepolo 11, 34143 Trieste, Italy\\ \email{andrea.biviano@inaf.it}
   \and
   IFPU-Institute for Fundamental Physics of the Universe, via Beirut 2, 34014 Trieste, Italy       
    }

   \date{Received \ldots; accepted \ldots}

  \abstract{The study of cluster substructures is important for the determination of the cluster dynamical status, assembly history, and the evolution of cluster galaxies, and it allows to set of constraints on the nature of dark matter and cosmological parameters.}{We present and test \texttt{DS+},
  a new method for the identification and characterization of group-sized substructures in clusters.}{Our new method is based on the projected positions and line-of-sight velocities of cluster galaxies, and it is an improvement and extension of the traditional method of \citet{DS88}. We test it on cluster-size cosmological halos extracted from the IllustrisTNG simulations, with virial masses $\rm{14 \lesssim \log (M_{200}/M_{\odot}) \lesssim 14.6}$, that contain on average $\sim 190$ galaxies.
  We also present an application of our method on a real data set, the Bullet cluster.}{\texttt{DS+} is able to identify $\sim 80\%$ of real group galaxies as members of substructures, and at least 60\% of the galaxies assigned to substructures belong to real groups. The physical properties of the real groups are significantly correlated with those of the corresponding detected substructures, albeit with significant scatter, and overestimated on average. Application of the \texttt{DS+} method to the Bullet cluster confirms the presence and main properties of the high-speed collision and identifies other substructures along the main cluster axis.}{\texttt{DS+} proves to be a reliable method for the identification of substructures in clusters. The method is made freely available to the community as a Python code.}


   \maketitle
%

\section{Introduction}\label{s:intro}
In the framework of the Cold Dark Matter cosmological model with cosmological constant ($\rm{\Lambda CDM}$) the assembly of dark matter halos proceeds hierarchically, i.e. small ones form first while larger ones form latter. Clusters of galaxies are the latest virialized structure to form through mergers of groups and individual galaxies. The accretion process of groups into clusters is revealed by the presence of substructures (or subclusters), which are secondary peaks in the distribution of galaxies, intra-cluster (IC) gas, and/or the cluster mass itself, on scales larger than the typical size of galaxies. 

The identification and characterization of cluster substructures is important in many ways. It allows to test the cosmological model of halo assembly \citep[e.g.,][]{RLT92,MEFG95,Thomas+98,SHYO03,PD07,FRGY10,ABK21}, to improve our understanding of the evolutionary mechanisms of galaxies in high-density regions \citep[e.g.,][]{Bekki99,Dubinski99,Gnedin99,Poggianti+04,TB08,Mahajan13,RLR13,Olave-Rojas+18,Bellhouse+22}, to constrain the nature of dark matter \citep[DM; see e.g.,][]{Markevitch+04,Clowe+06,Merten+11,Fischer+22}, and to identify clusters with an unrelaxed dynamical status caused by mergers, that can lead to biased estimates of 
the cluster mass  \citep[e.g.,][]{MHBN05,Biviano+06,VVDA08,TNM10,AB12,BGB13,Lagana+19,Zhang+22}. 

Thousands of clusters have been investigated for the presence of substructures, in several different ways \citep[e.g.][]{Miller+05,Lopes+06,Gal+09,WH13,SR19,Zenteno+20,Ghirardini+22,YHW22}. Despite this large statistics, the fraction of clusters embedded with substructures has been difficult to establish with precision. The most sensitive tests report fractions $\gtrsim 0.5$ \citep{KBPG01,Lopes+06,Ramella+07,WH13}, but this value depends very much on the method of detection and on the adopted significance level  \citep{KBPG01,Lopes+06}. 
Sample selection is also an issue to be considered when trying to establish the fraction of clusters with substructures. This fraction appears to increase with redshift \citep{APMG09,MJFVS09,Ghirardini+22}, and to be higher for
clusters detected by the Sunyaev-Zeldovich effect \citep{SZ69} than for clusters selected in the X-ray \citep[e.g.][]{Lopes+18,Campitiello+22}. Finally, the fraction of clusters with substructures is probably not a well-defined quantity, since there is a smooth transition between "regular" and "irregular" clusters \citep{DeLuca+21,Campitiello+22,Ghirardini+22}. 

Several methods exist for the detection of 
cluster substructures. Substructures can be -- and have been -- identified by the analysis of the projected phase-space distribution of cluster galaxies \citep[e.g.,][]{GB82,Pinkney+96,Einasto+12}, by the surface-brightness and temperature distribution of the X-ray emitting intra-cluster gas \citep[e.g.,][]{BHB92,Hashimoto+07,Zhang+09},
or by the presence of peaks in the maps of projected mass, derived using the gravitational lensing technique \citep[e.g.,][]{ASW98,Jauzac+16,Martinet+16}. The presence of cluster-scale radio halo emission and/or of wide-angle radio galaxies are also useful indicators of departure from dynamical relaxation \citep[but not always, see][]{WB13,BG18} and hence of the presence of major substructures \citep[e.g.,][]{Oklopcic+10,WH13,Wilber+19}.
Since the IC gas is a collisional component, and galaxies and DM are not, these different tracers often identify different substructures, and a full understanding of the cluster assembly history requires a multi-tracer approach \citep[e.g.,][]{Ferrari+05,GDRB05,CBS12,Ruppin+20}.

Most methods for substructure detection do not aim to identify the individual substructures but only to establish a cluster dynamical state. Useful indicators of a cluster dynamical state are the morphology of the IC gas surface brightness and/or the galaxy spatial projected distribution, measured by the concentration and the asymmetry parameter,  \citep[e.g.,][]{Pinkney+96,Lopes+06,Parekh+15,Bartalucci+19,Ghirardini+22}, or by more sophisticated techniques employing the 2D power spectrum of the IC gas or lensing mass distribution \citep{BT95,Mohammed+16,Campitiello+22}.
Other useful indicators are the offsets between the centroids of the various cluster components, galaxies, IC gas, DM \citep[e.g.,][]{Zenteno+20,DeLuca+21}.

Cluster morphology alone is not always a faithful indicator of its dynamical state \citep{SS21} and different morphological metrics do not always give a consistent picture on the cluster dynamical relaxation \citep{CBV21}. Additionally, very powerful information on a cluster dynamical state can come from the IC gas temperature distribution  \citep{Hashimoto+07,Zhang+09,Akamatsu+16,Lagana+19} and from the velocity distribution of cluster galaxies \citep{Muriel2002, Burgett+04,Miller+04,Ribeiro+13,Golovich+19,RP19,SR19,Sampaio+21}. A combination of the spatial and velocity distribution of cluster galaxies provides more powerful tests for the presence of substructures \citep[][and references therein]{DS88,CD96,GB02}.

A further step in the study of cluster substructures, beyond the general assessment of the cluster's dynamical state, is the identification of individual substructures. Their detection comes from the identification of peaks in the total projected mass \citep[as identified by weak lensing, see e.g.,][]{CGM04,Leonard+07,Jauzac+16,King+16}, from the identification of residuals in the X-ray cluster image after subtraction of a smooth model \citep[e.g.,][]{Neumann+03a,ASLNL12}, from X-ray temperature maps \citep[e.g.,][]{Zhang+09}, and from 2D maps of the density of galaxies in projected space \citep[e.g.,][]{Pisani96,Ramella+07,Girardi+11}, eventually complemented with the spectroscopic information \citep[e.g.,][]{Escalera+94,Girardi+15}.

Even more complicated is distinguishing which cluster galaxies belong to which substructures. In fact, tidal effects reduce the density of the infalling groups, whose size and internal velocity dispersion are doubled in $\sim 1$-3 Gyr since cluster infall \citep{BSA20}. Half of the infalling group galaxies escape the gravitational potential of the group after the first cluster pericenter passage and only those galaxies located very near the group center remain bound to it \citep{Haggar+22}.

There are only a few methods that allow identifying the galaxies that belong to substructures, \texttt{DEDICA} \citep{Pisani93,Pisani96}, \texttt{S-tree} \citet{GHK94}, the \texttt{h-method} \citep{SG96}, \citet{FR06}'s \texttt{mclust}, extensively used by \citet{Einasto+10,Einasto+18,Einasto+21}, 
\texttt{$\sigma$ plateau} \citep{YSDB15}, and \texttt{Blooming Tree} \citep{YDSB18}. Only for the latter two methods, a detailed assessment of their performances has been done using cluster-size halos in cosmological numerical simulations. In this paper, we introduce another method that allows to identify not only cluster substructures but also the galaxies that belong to them. It is an evolution of the classical method of \citet{DS88}, and we name it \texttt{DS+} after the authors' initials. It was already briefly introduced in the Appendix of \citet{Biviano+17a}. In this paper, we test the method using cluster-size halos extracted from cosmological hydrodynamical simulations and present a real-data application of the method itself.

The structure of this paper is the follows. We describe the method in Sect.~\ref{s:method}, and the numerical simulations in Sect.~\ref{s:sims}. We present the results of applying the method on the simulated halos in Sect.~\ref{s:res}.
In particular, in Sect.~\ref{ss:CP} we estimate the completeness and purity of the method, and in Sect.~\ref{ss:props} we compare several properties of \texttt{DS+} substructures with those of their corresponding real groups. In Sect.~\ref{s:bullet}
we apply the \texttt{DS+} method to a real data set \citep[the Bullet cluster,][]{Markevitch+02}, and we give our summary and conclusions in Sect.~\ref{s:conc}.

\section{The \texttt{DS+} method} \label{s:method}
The original method on which \texttt{DS+} is based, was developed by \citet{DS88}. They estimated the differences, $\delta$, between the mean velocities and velocity dispersions of the whole cluster and all possible substructures defined by $N_g=11$ neighboring cluster galaxies  \citep[see Eq.~(1) in][]{DS88}. When there are strong deviations of the local galaxy velocity field from the global one, the sum of these $\delta$ differences, $\Delta$, divided by the whole cluster velocity dispersion, becomes much larger than the number of cluster members, $N_m$, and the cluster is likely to contain substructures. The likelihood is evaluated via a Monte Carlo technique in which cluster galaxy velocities are randomly shuffled with respect to their coordinates,  to erase any possibly existing spatial-velocity correlations. In the original implementation, this method does not identify the substructures, nor the galaxies in substructures, it only provides a global probability for the cluster to be in an unrelaxed dynamical state because of the presence of substructures.
 
The original method has evolved with time.  \citet{Bird94} used $N_g = N_m^{1/2}$, instead of the very {\em ad-hoc} value of 11. In comparing the velocity dispersions of the whole cluster and the candidate substructures, \citet{Biviano+02} discarded as not significant the cases in which the substructure velocity dispersion turned out to be larger than the cluster one. The rationale behind this choice is that velocity dispersion is a mass proxy and groups must be less massive than the cluster they are falling into. In addition, \citet{Biviano+02} considered the full distribution of the $N_m$ $\delta$ values,
rather than just their sum. By comparing the observed $\delta$ distribution with Monte Carlo realizations obtained by azimuthally scrambling the galaxy positions, the authors estimated the probability for a given $\delta$ value to be significantly larger than the average of cluster members. As a result, they could identify which galaxies have the highest probability of belonging to substructures, but they would not identify the substructures themselves.

Rather than estimating the value of $\delta$ from the combined difference in mean velocity and velocity dispersion,  \citet{Ferrari+03} separated the two contributions, $\delta_v$ and $\delta_{\sigma}$, respectively. \citet{Girardi+15} went beyond the implicit isothermal assumption of the original method (a cluster with constant velocity dispersion at all radii) and instead of using the whole cluster velocity dispersion in the estimation of $\delta$, they used the cluster velocity dispersion profile.

The \texttt{DS+} method includes all these previous modifications of the original test of \citet{DS88}, and it introduces significant new features.
Possible substructures are considered around each cluster member, but we do not enforce a given number of substructure members. We consider substructures of several possible multiplicities, $N_g(j)=j$, $j=3, \ldots, k$, where $k$ is the smallest value of
$j$ for which $N_g(k)> N_m/3$. In doing this we effectively take into account that substructures of different richness coexist in a given cluster and that the largest substructures we consider can contain more than 1/3 of all cluster galaxies\footnote{Extending the maximum to $N_m/2$ would create an ambiguity about which is the cluster and which is the substructure.}.

We define $\delta_v$ and  $\delta_{\sigma}$ as in \citet{Biviano+02},
\begin{equation}
\delta_v= N_g^{1/2} \, \mid \overline{v_g} \mid \, [(t_{n}-1) \, \slos(R_g)]^{-1},
\end{equation}
and
\begin{equation}
\delta_{\sigma}=[1-\slosg/\slos(R_g)] \, \{1-[(N_g-1)/\chi^+_{N_g-1}]^{1/2}\}^{-1},
\end{equation}
where $R_g$ is the average projected substructure distance
from the cluster center, $\overline{v_g}$ is the mean substructure velocity, $\slos(R)$ is the cluster line-of-sight (l.o.s. hereafter) velocity dispersion profile, and $\slosg$ is the substructure l.o.s. velocity dispersion (galaxy velocities are in the cluster rest-frame). Following \citet{Biviano+02} only positive values of $\delta_{\sigma}$ are considered, that is, group velocity dispersions that are higher than the cluster one, are not considered to be significant. However, \texttt{DS+} can still identify substructures with velocity dispersions larger than the cluster if they are characterized by a large $\delta_v$ value.

The Student-$t$ and $\chi^2$ distributions are used to normalize the differences in units of the uncertainties in the mean velocity and velocity dispersion, respectively \citep[see][]{BFG90}. We assume a null cluster mean velocity at all radii, that is, there is no cluster rotation \citep[the fraction of clusters with evidence for rotation is $\lesssim 1$\%, see][]{HL07}. We use the biweight estimator for
$\slosg$ and $\slos$ for samples of 15 galaxies or more, and the gapper estimator for smaller samples \citet{BFG90}. 

The cluster line-of-sight velocity dispersion profile, $\slos(R)$, can be directly estimated from the cluster member
velocities, using the \texttt{LOWESS} smoothing algorithm
\citep{Gebhardt+94}. In alternative, $\slos(R)$ can be estimated
by assuming a theoretical model. We adopt the NFW model for the cluster mass profile \citep{NFW97}, with a total mass obtained from $\slos$ via a scaling relation \citep{MM07} and a concentration given by the relation of \citet{MDvdB08}. We adopt the velocity anisotropy profile of \citet{MBM10}. The cluster $\slos(R)$ is obtained by applying the Jeans equation of dynamical equilibrium and the Abel projection equation
\citep[Eqs.~(8), (9), and (26) in][]{MBB13}. 

We estimate the probability of $\delta_v$ and $\delta_{\sigma}$ by comparing them with the corresponding values obtained for a suitable number (typically 500) of
MonteCarlo resamplings in which we replace all the cluster
galaxy velocities with random Gaussian draws from a distribution of zero mean and dispersion equal to $\slos(R_g)$. We consider as statistically significant those
substructures with $\delta_v$ and/or $\delta_{\sigma}$ value probabilities $\leq 0.01$.
\begin{figure*}
\centering
\includegraphics[width=\hsize]{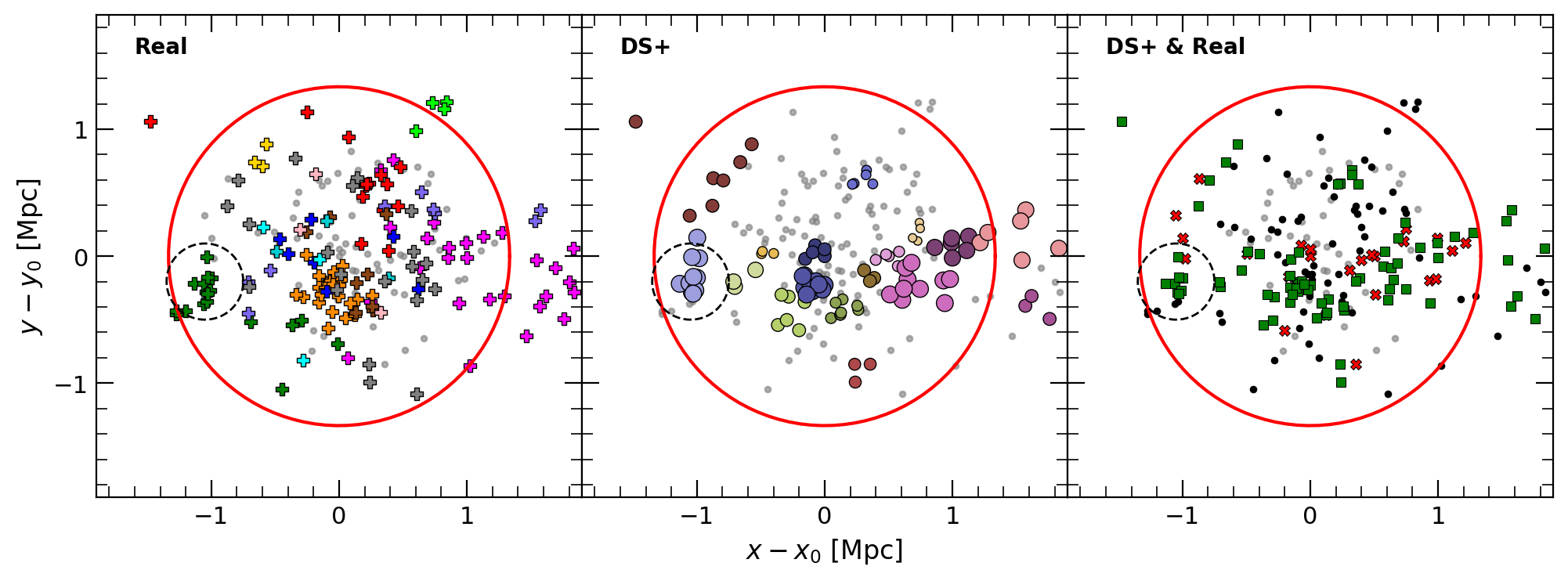}
  \caption{Projected spatial distribution of galaxies in a simulated TNG100 cluster at $z=0$, of $\rm{r_{200} = 1.3~Mpc}$ and $\rm{M_{200} = 2.5 \times 10^{14}~ M_{\odot}}$. In all panels, coordinates are in Mpc from the cluster center, defined as the position of the particle with the minimum gravitational potential energy,
  the red circle represents the virial radius $\rm{r_{200}}$, and the black dashed-line circle highlights the position of a real group of galaxies, that we discuss in the text.
 \textit{Left:} Crosses correspond to the galaxies that were accreted as part of groups; grey dots identify the galaxies that entered the cluster individually. \textit{Center:} Circles of different colors identify galaxies assigned to different substructures by \texttt{DS+} in its no-overlapping mode, the circle sizes being proportional to the individual probability of each DS+ group. Grey dots represent galaxies unassigned to any substructure. \textit{Right:} Green squares identify galaxies in real groups that are correctly assigned to substructures by \texttt{DS+} method, red crosses indicate galaxies that entered the cluster alone, that are incorrectly assigned to substructures, and small black dots identify galaxies that were accreted as part of real groups but were not assigned to substructures by \texttt{DS+}.} 
     \label{f:distribution}
\end{figure*}

At this stage of the method, the statistically significant substructures may be overlapping, that is two significant substructures could share one or more galaxies. 
If the final aim of the method is to identify which galaxies belong to substructures, the \texttt{DS+} code can be stopped here. We call this the "overlapping" mode of \texttt{DS+}. On the other hand, if the final aim of the method is to identify the individual groups that are falling or have fallen into the cluster, we must continue the procedure in what we call the "no-overlapping" mode of \texttt{DS+}. To ensure that the substructures are uniquely defined, that is, that a given galaxy is not assigned to more than one substructure, we proceed as follows. If a given galaxy is assigned to more than one significant substructure, we assign it to the most significant one, that is the one with the lowest $\delta_v$ and/or $\delta_{\sigma}$ probability. All the other substructures containing this galaxy are then removed from the list of significant groups. 

Finally, in the "no-overlapping" mode of \texttt{DS+}, we adopt a method to address the problem of fragmentation, that is when two or more substructures are fragments of larger physical groups. We merge two substructures if their extents in l.o.s. velocity and projected spatial distance are larger than their mean velocity difference and the separation between their centers, respectively, that is we require the following conditions to apply:
\begin{equation}
     d_{i,j} < {\rm max}(d_{\rm{max},i},d_{{\rm max},j}) \, \wedge \,
    \mid \overline{v_{g,i}}-\overline{v_{g,j}} \mid < {\rm max}(\mid v_{{\rm max},i} \mid, \mid v_{{\rm max},j} \mid).
    \label{eq:merge}
\end{equation}
In Eq.~(\ref{eq:merge}) $d_{i,j}$ is the projected distance between the median centers of groups $i$ and $j$, and $d_{{\rm max},i}$ is the maximum distance of any galaxy of the group $i$ from its group center, $\overline{v_{g,i}}$ is the mean l.o.s. velocity of the group $i$, and $\mid v_{{\rm max},i} \mid$ is the maximum absolute velocity difference of any galaxy of group $i$ from its group mean velocity. 

The \texttt{DS+} method has been coded in \texttt{MilaDS}, developed in Python 3, and is freely available for use at a GitHub\footnote{ \href{https://github.com/josegit88/MilaDS}{https://github.com/josegit88/MilaDS}} repository.

\section{Numerical simulations}
\label{s:sims}

We test the \texttt{DS+}method using The Next Generation Illustris Simulations \citep[IllustrisTNG \footnote{ \href{https://www.tng-project.org/}{https://www.tng-project.org/}}, ][]{Pillepich2018a, Pillepich2018b, Springel2018, Nelson2019TNG}, a suite of $\rm{\Lambda}$CDM magneto-hydrodynamic cosmological galaxy formation simulations. IllustrisTNG is an improved version of its predecessor Illustris \citep{Vogelsberger2014a, Vogelsberger2014b} with improved physical models, and comes in boxes of different sizes and resolution per particle (known as TNG50, TNG100, and TNG300) that allow studying the formation and evolution of galaxies on different scales and several environments.

In particular, for this work we use data from IllustrisTNG100-1 (TNG100 hereafter) which corresponds to a periodic cosmological box of 110.7 Mpc side and resolution per particle of $\rm{m_{dm} = 7.5 \times 10^6 ~ M_{\odot}}$ for DM and $\rm{m_{gas} = 1.4 \times 10^6 ~ M_{\odot}}$ for gas cells, with a softening-length of 0.74 kpc (at redshift $z=0$), although the hydrodynamics can reach a higher spatial resolution in the high-density regions. 
The simulation is performed using the moving mesh
\texttt{AREPO} code \citep{Springel2010}, and there are also subgrid physics details. The initial conditions of the simulation were established at $z = 127$ using Zeldovich’s approximation and the \texttt{N-GENIC} code \citep{SpringelNGENIC} and cosmological parameters consistent with results from the \citet{PlankColaboration2016}: $\Omega_m = \Omega_{\rm{dm}} + \Omega_{bar} = 0.3089$, cosmological constant $\Omega_{\Lambda} = 0.6911$, with $h=0.6774$ and $\sigma_8 = 0.8159$. The identification of the halos and subhalos is done using \texttt{Friends-of-Friends} \citep[FoF,][]{Davis1985} and \texttt{SUBFIND} \citep{Springel2010}. To follow halos and subhalos over time we used the \texttt{SUBLINK merger-trees} \citep{Rodriguez-Gomez+15}. 

In this paper, we select the fourteen most massive halos included in the TNG100 box at $z = 0$, corresponding to galaxy clusters with virial mass $\rm{M_{200} \gtrsim 10^{14} ~ M_{\odot}}$. Within these 14 host halos, we consider all galaxies with stellar mass $\rm{M_{\star} \geq 1.5 \times 10^8 ~ M_{\odot}}$, corresponding to an average of $\sim 120$ stellar particles in the lowest mass objects. On average, there are 190 galaxies per halo (from $\sim$ 300 galaxies in the most massive halo to $\sim$ 100 in the least massive). Afterwards, we follow their time evolution to obtain information on which galaxies fell as individual objects or as part of groups, as was done in \citet{BSA20}.\\

\begin{figure*}
\centering
\includegraphics[width=\hsize]{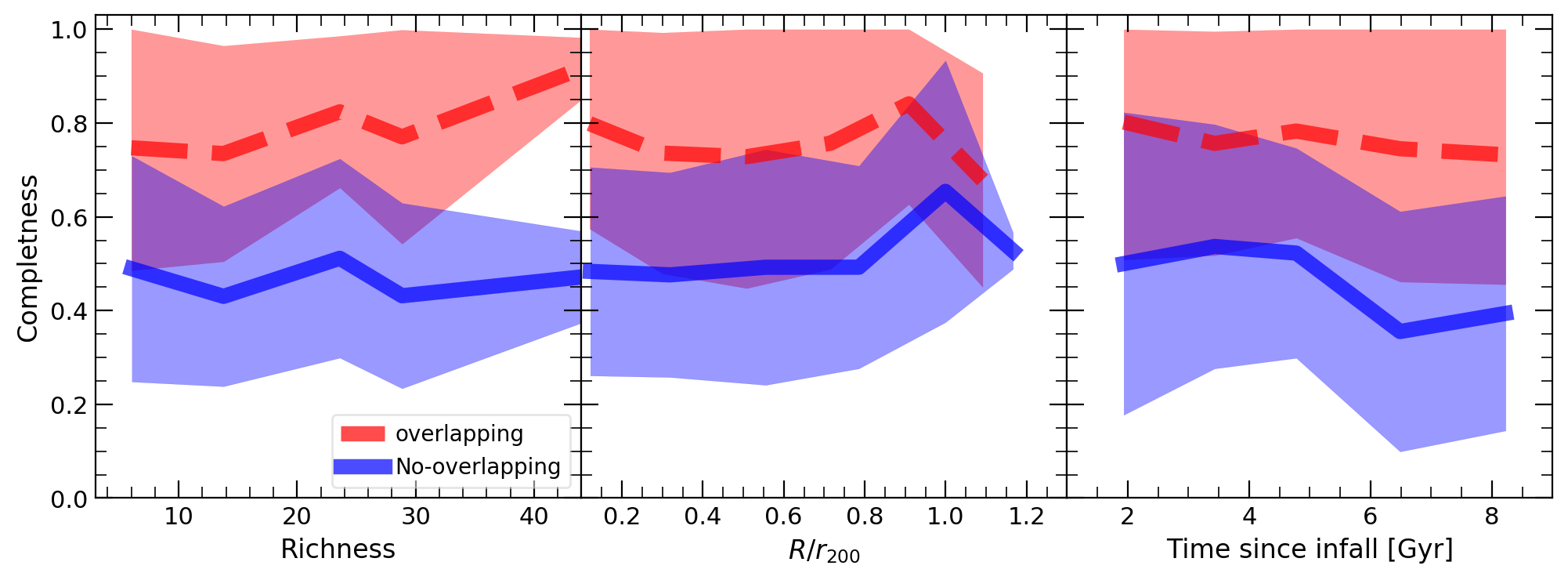}
  \caption{Completeness $C$ of member galaxies of real groups in simulated TNG100 clusters detected by \texttt{DS+} groups. In all panels, the red dashed line indicate the average over all clusters, using the overlapping mode, while the blue solid line refers to the no-overlapping mode. The filled areas indicate one standard deviation. In all cases, the curves correspond to the stacking of all analyzed clusters. \textit{Left:} $C$ as a function of the richness of the real groups, corresponding to the number of galaxies detected in substructures.  \textit{Center:} $C$ as a function of the real group cluster-centric distance. \textit{Right:} $C$ as a function of the time since group infall.}   
     \label{f:completeness}
\end{figure*}

\begin{figure*}
\centering
\includegraphics[width=\hsize]{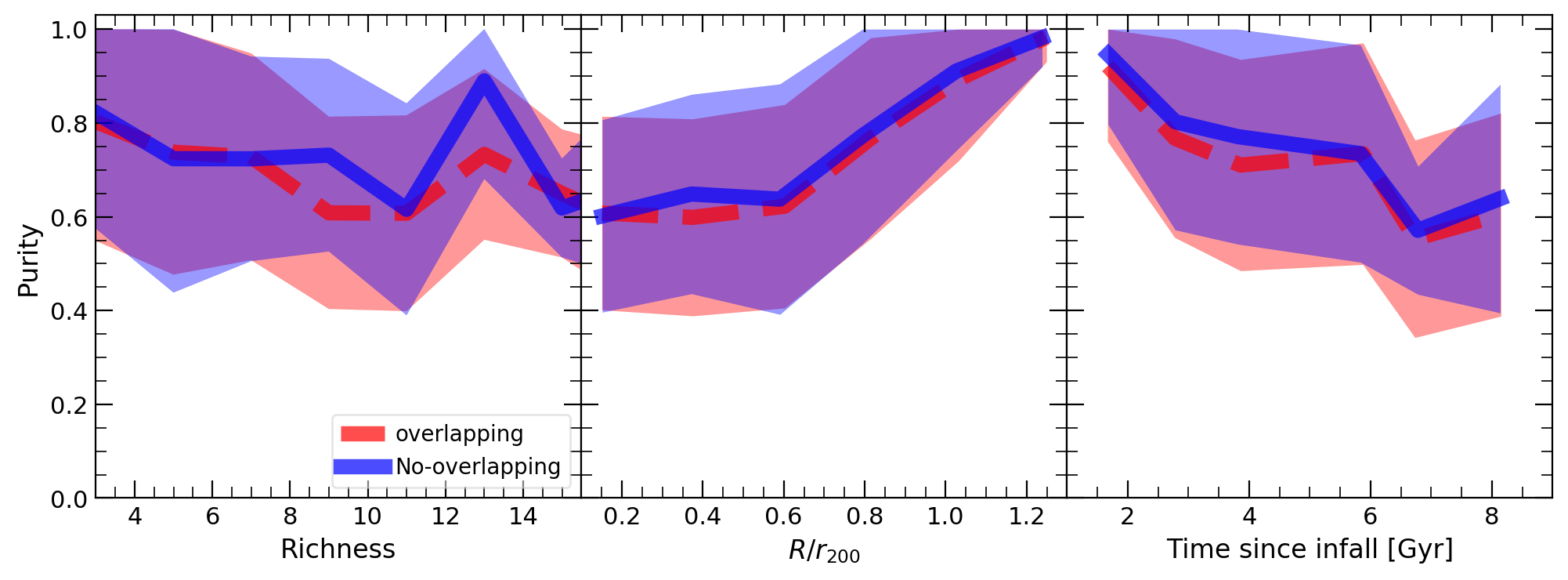}
  \caption{Purity $P$ of \texttt{DS+} detected substructures in simulated TNG100 clusters. Lines and colors have the same meaning as in Fig.~\ref{f:completeness}. The quantities on the x-axis of the three panels are the same as in Fig.~\ref{f:completeness}.} 
     \label{f:purity}
\end{figure*}

\section{Testing \texttt{DS+} on simulated clusters} 
\label{s:res}

We applied the \texttt{DS+} method to the 14 simulated halos observed at the present time, considering the information about the infall of their groups since $\sim 8$ Gyr ago.
We considered each of the three orthogonal projections like an individual cluster, for a total of 42 galaxy clusters used in our analysis. Our aim is to identify those galaxies that entered the cluster in groups of at least three members, using only projected coordinates and l.o.s. velocities. Hereafter we use the term ``real groups'' to refer to the galaxies that were part of groups identified in the simulations at high redshift (before the infall) and the term "substructures" for the groups identified by \texttt{DS+}, using the information of l.o.s. velocities and 2D spatial coordinates (at a redshift of interest).

In Fig.~\ref{f:distribution} we show one cluster at the last time in the simulation, distinguishing between galaxies that entered the cluster as individuals and galaxies that entered the cluster in groups. A large fraction of the cluster galaxies were accreted in groups \citep[$\sim 60 \%$, see][]{BSA20} and it is certainly impossible to identify all of them as substructures as many are well mixed with the cluster galaxies that entered the cluster individually. In the central panel of Fig.~\ref{f:distribution} we show the substructures and in the right panel we show the comparison between the real groups and the substructures identified by our method. Another example was added in Fig.\ref{f:distribution_app} in the appendix, for the evolution of the same cluster, in this figure we present similar information for each $\sim 2$ Gyr in look-back time.

\subsection{Completeness and Purity} \label{ss:CP}
A more general assessment of the performance of our method can be gained by evaluating the completeness and purity of the samples of galaxies in real groups and substructures, respectively. We call completeness, $C$, the fraction of galaxies in real groups that are also detected as members of any substructure
\begin{equation}
     C = \frac{N_{\rm{DS+}}}{N_{\rm real}} \ ,
    \label{eq:complet}
\end{equation}
and purity, $P$, the fraction of galaxies in detected substructures that belong to any real group
\begin{equation}
     P = \frac{N_{\rm{DS+,real}}}{N_{\rm{DS+}}}.
    \label{eq:purity}
\end{equation}

In Fig.~\ref{f:completeness} we show $C$ as a function of different variables: richness, i.e., the number of galaxies of the real groups detected as members of the \texttt{DS+} groups, the $2D$ projected distance from the center of the cluster in units of the virial radius\footnote{The virial radius of the cluster $r_{200}$ is the radius of a sphere with an over-density 200 times the critical density of the Universe.} ($R / r_{200}$), and the time since group infall into the cluster\footnote{In this work, we follow the infall time definition of \citet{BSA20}, that is the last time the infalling group and the cluster were identified as different FoF systems.}. We see that $C \sim 0.8$ for the overlapping mode of \texttt{DS+}. $C$ is lower ($\sim 0.5$) for the no-overlapping mode, as expected given that in this mode we discard all substructures that have galaxies in common with more significant substructures. The fact that $C$ values close to 1 are not observed (in particular in the non-overlapping mode), corresponds to the fact that the actual groups that fell into a cluster tend to disperse significantly after the first pericentric passage \citep{Choque-Challap2019, BSA20,Haggar+22}. 

$C$ does not show a strong dependence on group richness, projected cluster-centric distance, or time since infall. For the overlapping mode, $C$ increases with the richness only mildly, reaching a value of $\sim 0.9$ for groups of $\sim 40$ members. In the no-overlapping mode, $C$ mildly increases with group distance from the cluster center. The increasing trends of $C$ with group richness and projected cluster-centric distance are expected since richer groups offer better statistics for detection, and at larger cluster-centric distances the density contrast of the groups is larger relative to the cluster.

The dependence of $C$ on the time since infall is less strong than would be expected from the fact that infalling groups double their size $\sim 1 - 3$ Gyr after infall \citep[depending on the group-mass,][]{BSA20}, and in many cases are completely destroyed after their first passage through the pericenter. However, the collisionless nature of the group galaxies allows them to retain a mostly consistent velocity even after pericenter passage. This allows them to be identified as members of substructures (even if not in a single one) allowing $C$ not to drop too rapidly with time since infall. 

As an example, the substructure represented by the dark green crosses in the left panel of Fig.~\ref{f:distribution} (highlighted with the dashed black circle) corresponds to a substructure detected by \texttt{DS+} and indicated by light purple dots in the middle panel of the same figure. However, in other cases, real group galaxies are not associated to any substructure (small black circles in the right panel), or they are associated to many substructures and not, in major part, to a single one (e.g., the group represented by the magenta crosses in the left panel). This occurs because the group has already crossed the center of the cluster, experiencing strong tidal forces that deform the shape of the primitive association.

In Fig.~\ref{f:purity} we show $P$ as a function of the same variables as in Fig.~\ref{f:completeness}. There is no strong dependence of $P$ on the mode of operation (with or without overlapping), which shows that when we run the \texttt{DS+} method in the overlapping mode, we do not add substantial noise to the purity result of the detected \texttt{DS+} groups. This is important for the estimate of the properties of group galaxies.

$P$ is above $60 \%$ and can approach $100\%$ in some cases. It decreases as group richness increases, from $80\%$ to just over $60\%$. However, this slight decreasing trend does not seem very significant, so the purity could be considered more or less flat, around $\sim 70\%$ for any multiplicity.

$P$ increases significantly with the projected distance from the cluster center to the outskirts, from $\sim 60$\% near the cluster center to almost 100\% beyond the virial radius. It is to be expected that at large distances from the cluster center (where the cluster density is sufficiently low) the contamination of the substructure by cluster members that are not in groups would be less significant. Many of these \texttt{DS+} substructures would correspond to recent accretions or to fragments of real groups that are close to their first apocenter. $P$ also shows a clear decreasing trend with time since infall. This is probably due to the fact that when a group crosses the cluster its size increases considerably by tidal effects, allowing more interlopers to contaminate the region occupied by the group in projection.

The results presented above were obtained using an upper probability limit of 0.01. Similar values of $C$ and $P$ were obtained when considering lower probability limits (e.g., 0.005), although of course with fewer \texttt{DS+} substructures.

It is interesting to briefly compare our results with those obtained for the recently developed \texttt{Blooming Tree}, which has been claimed to be the best substructure identification method, and superior to \texttt{$\sigma$ plateau} \citep{YDSB18}. A direct comparison is not possible, because of the different simulations used, and the different definitions of completeness and purity \citep["success rate" in][]{YDSB18}. Summarizing from \citet{YDSB18}'s results, \texttt{Blooming Tree} reaches a completeness $C \sim 0.8$, and a purity, $P \sim 0.6$, for $\sim 1/2$ the detected structures. The completeness of \texttt{Blooming Tree} is therefore comparable to that of \texttt{DS+} in its overlapping mode, and superior to that of our method in its no-overlapping mode. The purity of \texttt{DS+} substructures appears to be superior to that of  \texttt{Blooming Tree}, since the value $P=0.6$ is reached by the latter method only for $\sim 1/2$ of the detected structures. Pending a more direct comparison between \texttt{Blooming Tree} and \texttt{DS+}, which is beyond the scope of this paper, we tentatively conclude that these two algorithms reach similar performances in the detection of substructures.

\begin{figure*}
\centering
\includegraphics[width=\hsize]{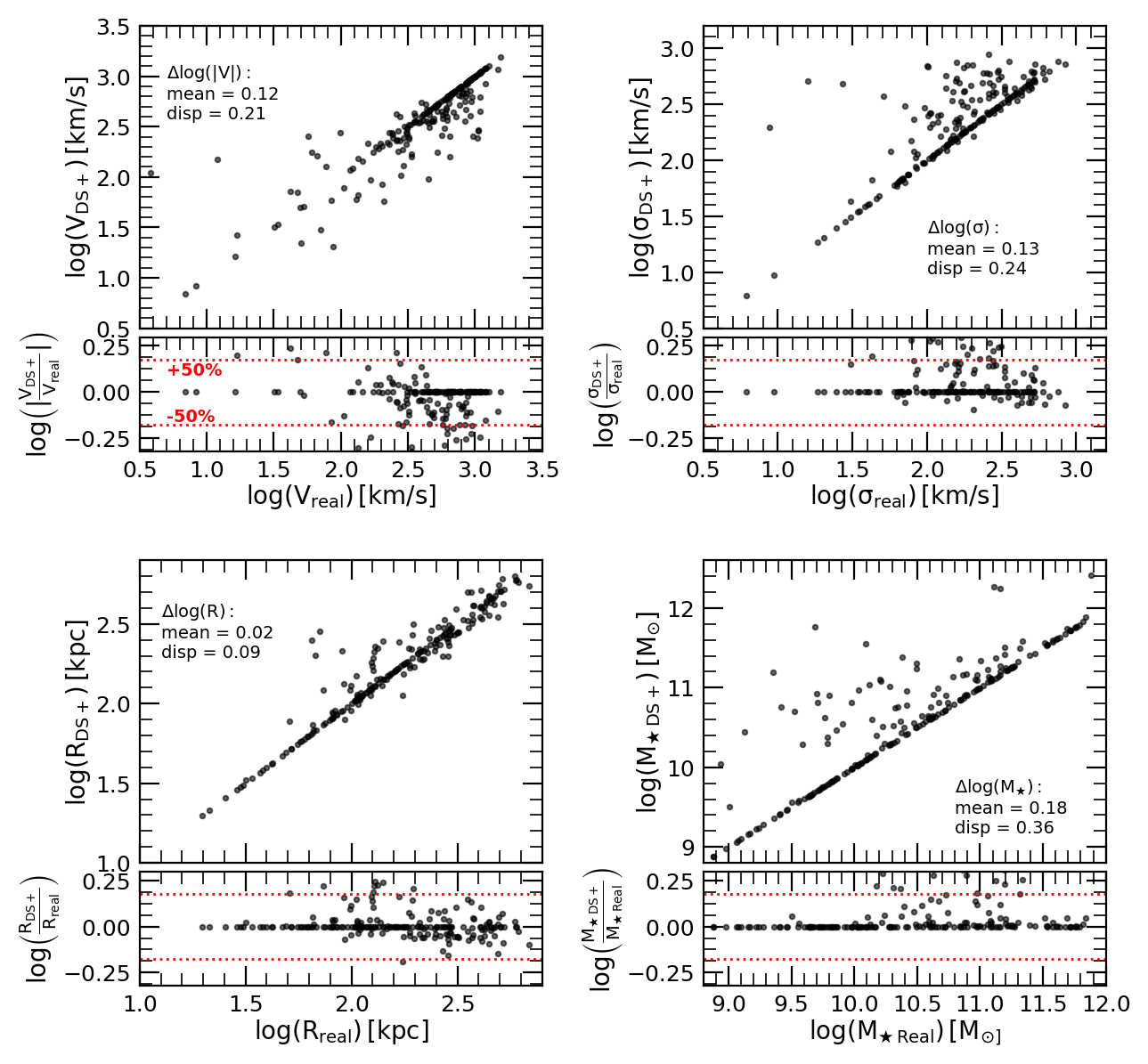}
  \caption{Comparison of different properties of the \texttt{DS+} no-overlapping substructures with the corresponding ones of the real groups matched to the detected substructures. \textit{Top left:} mean l.o.s. velocities. \textit{Top right:} l.o.s. velocity dispersions. \textit{Bottom left:} projected group size (mean harmonic radius). \textit{Bottom right:} total stellar mass. In each panel, we include the mean and standard deviation of the differences between the substructure and the matched real group properties. The lower sub-panels show the logarithm of the ratio of the substructure and group properties, and the red dotted lines indicate 50\% variations with respect to a ratio of unity.} 
     \label{f:properties_comp}
\end{figure*}

\subsection{\texttt{DS+} substructures vs. real groups internal properties} 
\label{ss:props}

In order to have an additional estimate of the characteristics and reliability of the groups detected using the \texttt{DS+} method, we compare several of the global properties of the detected substructures with those of their corresponding real groups, as detailed below. The global properties we consider are the l.o.s. mean velocity and velocity dispersion, the $2D$ size of the group (as measured by the harmonic mean radius), and the total stellar mass.

The properties of each \texttt{DS+} substructure are evaluated using all galaxies assigned to that substructure. Since substructure members can be members of more than one real group, we only consider the group that has the largest number of members in common with the considered substructure. We then compute the properties of this real group using only its members that are also members of the substructure. This is done because many of its original group members are rapidly dispersed into the cluster after infall and attributing them to the group would not be correct in physical terms. 

The mean values of the ratios of the substructure and group properties are given in Table~\ref{t:table1}. In all cases, the ratios are above unity, but with considerable dispersions. The \texttt{DS+} estimates of both the velocity dispersion and the stellar mass of the groups are strongly overestimated, as expected because of the presence of interlopers in the substructures.

In Fig.~\ref{f:properties_comp} we show the correlations between the properties of the substructures and those of the corresponding real groups. In each panel we include the mean values of the property differences $\Delta \log(X) = \log(X_{DS+}) - \log(X_{Real})$ and their standard deviation. According to the Spearman correlation coefficient, all correlations are
significant with $0.94$ to $0.99$ probabilities (the lowest value is for the velocity dispersion). These values indicate that we can use the properties of the detected \texttt{DS+} substructures to predict the properties of the corresponding real groups. However, this is only true on average, as the inferred properties may be very different from the real ones for individual groups. Galaxy properties such as stellar population, metallicity, etc., could be used to identify and remove group interlopers to improve the correspondence between group and substructure global properties. However, we have not explored this possibility in this analysis.

\begin{table}[!t]
\centering
\caption{Mean ratios of the properties of \texttt{DS+} substructures and of real groups.}
\begin{tabular}{cccc}
\hline\hline\noalign{\smallskip}
\!\!\!\!mean Vel & \!\!\!\!Vel disp. & \!\!Size & \!\!\!\!Stellar Mass\!\!\!\!\\
\!\!\!\! $\rm{|V_{DS+} / V_{real} |}$ & \!\!\!\! $\rm{\sigma_{DS+} / \sigma_{real}}$ & \!\! $\rm{R_{DS+} / R_{real}}$ & \!\!\!\! $\rm{M_{\star, DS+} / M_{\star, real}}$ \!\!\!\!\\
\hline\noalign{\smallskip}
$1.2 \pm 2.3$ &  $1.8 \pm 3.0$ &\!\!$1.1 \pm 0.4$ &  $1.5 \pm 1.4$\\
\hline
\end{tabular}
\label{t:table1}
\end{table}

\section{An application: the Bullet Cluster} \label{s:bullet}
As a practical example of our \texttt{DS+} method, we apply it to the famous "Bullet" cluster 1E 0657-558 \citep{Barrena+02,Markevitch+02,CGM04}. We collect spectroscopic data for galaxies in the cluster region from the NASA/IPAC Extragalactic Database (NED\footnote{The NASA/IPAC Extragalactic Database (NED) is funded by the National Aeronautics and Space Administration and operated by the California Institute of Technology.}). After removing double entries, we find 231 galaxies with redshifts within a circle of 10$\arcmin$ radius around the cluster center. All these galaxies are members of the cluster, according to the Clean procedure of \citet{MBB13}. The mean cluster redshift is $\overline{z}=0.2965 \pm 0.0003$, and the rest-frame velocity dispersion is $\sigma_v=1163_{-59}^{+56}$~km~s$^{-1}$. These values are obtained using the biweight estimator as recommended by \citet{BFG90} for "large" data sets. These values are consistent with those determined by \citet{Barrena+02} using 78 cluster members.
\renewcommand{\arraystretch}{1.5}


\begin{table*}[!t]
\centering
\caption{Properties of detected Bullet cluster substructures}
\begin{tabular}{lrrrrrrr}
\hline\noalign{\smallskip}
ID & $N_g$ & $p$ &   $x$    &   $y$ &     $R/r_{200}$   &  $v_g$    & $\sigma_g$ \\
   &           &     &  [kpc]   &   [kpc] &    & [km~s$^{-1}$] &  [km~s$^{-1}$] \\
\hline
 1 &  6 & 0.000 &  -239 &   383 &   0.21 & $ -1070 \pm  694 $  & $ 1535_{-407}^{+537} $ \\
 2 &  7 & 0.000 &   443 &  -820 &   0.44 & $   383 \pm  106 $  & $  750_{-184}^{+238} $ \\
 3 &  7 & 0.000 &   837 &   294 &   0.42 & $  1113 \pm  547 $  & $ 1356_{-333}^{+431} $ \\
 4 &  9 & 0.000 &    -2 &   110 &   0.05 & $    12 \pm  264 $  & $  915_{-199}^{+250} $ \\
 5 &  6 & 0.001 &  -656 &  -589 &   0.41 & $   208 \pm  521 $  & $  978_{-259}^{+342} $ \\
 6 (Bullet) &  9 & 0.001 &  -736 &   223 &   0.36 & $   413 \pm  124 $  & $  849_{-185}^{+232} $ \\
 7 &  6 & 0.002 &    47 &  -424 &   0.20 & $ -1020 \pm  135 $  & $ 1260_{-334}^{+440} $ \\
 8 &  7 & 0.003 &    82 &  -105 &   0.06 & $  1105 \pm  614 $  & $ 1536_{-377}^{+488} $ \\
 9 & 11 & 0.008 & -1431 &   767 &   0.76 & $  -505 \pm  204 $  & $ 1131_{-223}^{+275} $ \\
10 &  9 & 0.010 &   173 &   264 &   0.15 & $  -163 \pm  255 $  & $  708_{-154}^{+193} $ \\
\hline\noalign{\smallskip}
\end{tabular}
\tablefoot{$N_g$ is the number of galaxies assigned to the substructure by the \texttt{DS+} algorithm. $x, y$ are the positions in kpc from the cluster center, as in Fig.~\ref{f:bullet_xyv}. $R$ is the distance in kpc from the cluster center. $v_g$ and $\sigma_g$ are the mean velocity and velocity dispersion, respectively, evaluated using the biweight and the gapper estimator, respectively. Errors are 1 $\sigma$.
Substructure no. 6, denoted by an asterisk, corresponds to the group giving the nickname to the Bullet cluster.}
\label{t:Bgroups}
\end{table*}

\begin{figure*}
\centering
\includegraphics[width=\hsize]{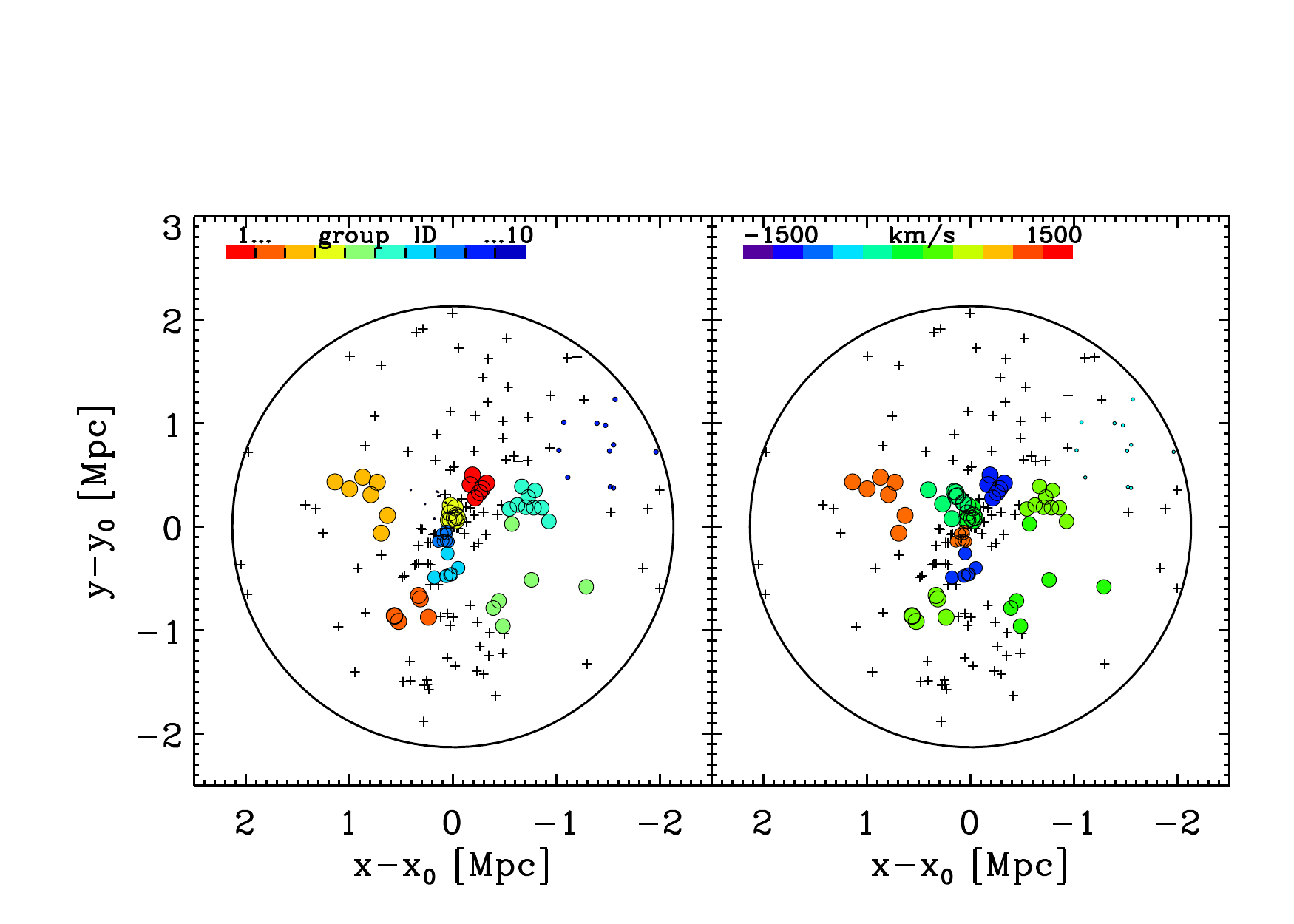}
\caption{Projected spatial distribution of Bullet member galaxies. North is up, East is to the left. The large black circle has a radius of $r_{200}=2.13$ Mpc and is centered on the cluster center, RA=104.65139, Dec=-55.95468. Smaller circles (resp. crosses) represent galaxies assigned (not assigned) to substructures by \texttt{DS+}. The size of the circles scales as $1-100 \, p$, where $p$ is
the probability of the detected group listed in Table~\ref{t:Bgroups}.  
{\it Left panel:} Different colors identify galaxies assigned to different groups, numbered 1 to 10 as in the inset bar and Table~\ref{t:Bgroups}. Group no. 6 is the Bullet,
represented by the nine turquoise dots at coordinates $(-0.74,0.22)$.
{\it Right panel:}
The color scale represents the mean velocity of the groups (see Table~\ref{t:Bgroups}). }
    \label{f:bullet_xyv}
\end{figure*}

\begin{figure}
\centering\includegraphics[width=\hsize]{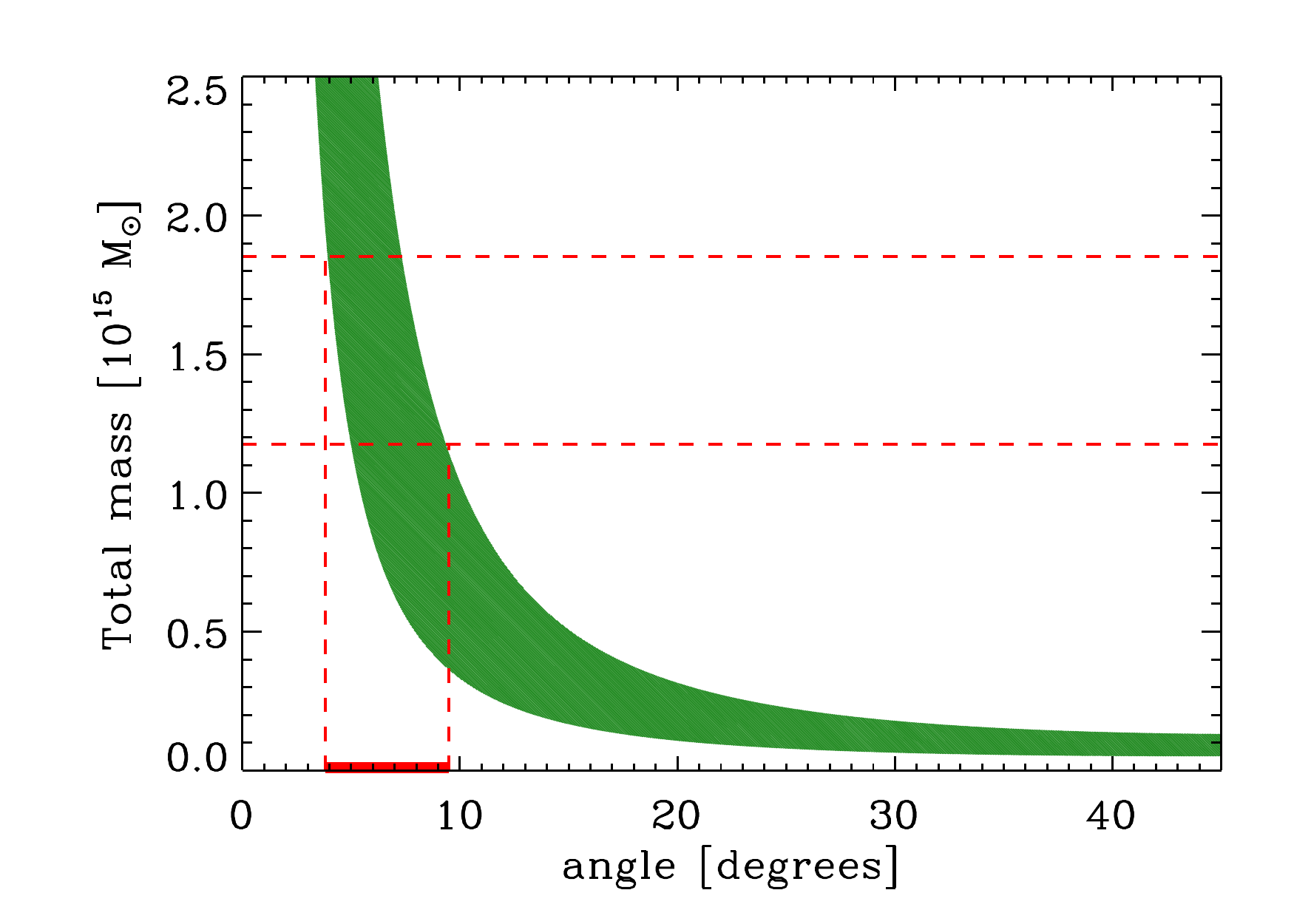}
\caption{Two-body collision model between the main cluster and the Bullet (group no. 6 in Table~\ref{t:Bgroups}). The total mass of the system is shown on the y-axis, as a function of the angle of the collision axis with respect to the plane of the sky. The estimated total mass range (1 $\sigma$) is illustrated by the two dashed lines. 
At the intersection of these lines with the model curve, we draw two vertical lines that identify the inferred allowed collision angles  (in green on the x-axis), $\sim 4^{\circ}$-$10^{\circ}$.}
\label{f:2body}
\end{figure}

\begin{figure}
\centering\includegraphics[width=\hsize]{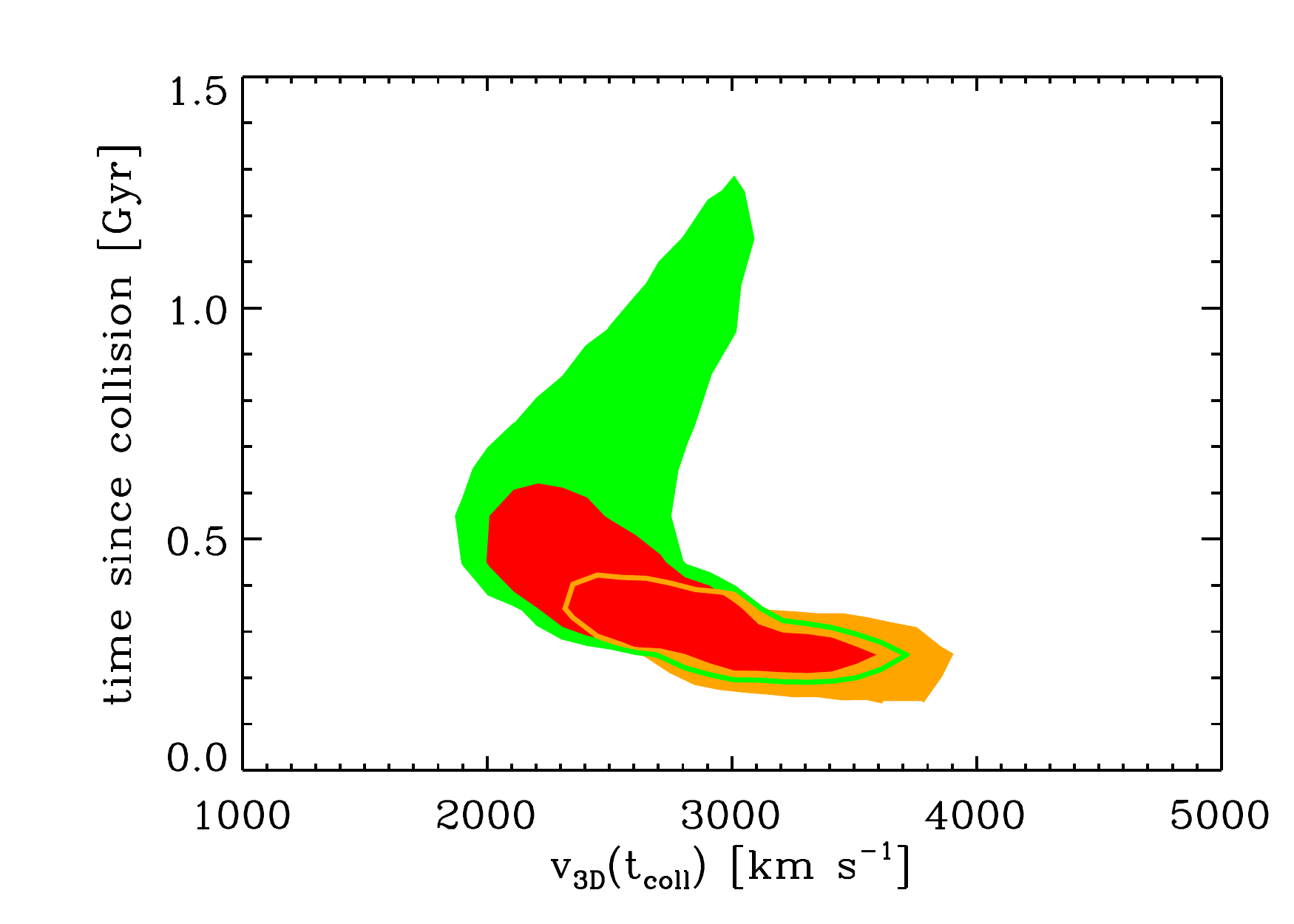}
\caption{Result of the \texttt{MCMAC} algorithm \citep{Dawson13} applied to the Bullet cluster and its bullet (group no. 6 in Table~\ref{t:Bgroups}), time since collision versus 3D velocity at the time of the collision. Contours are 68\% confidence levels. Green contours do not include any constraint on the collision angle. Red contours are obtained by considering only angles $\leq 29^{\circ}$, that is the 1 $\sigma$ constraint derived by \citep{WCN18}. Orange contours are obtained by considering only angles between $3^{\circ}$ and $10^{\circ}$, as inferred from the classical two-body collision model.}
\label{f:mcmac}
\end{figure}

Given $\sigma_v$, we estimate the Bullet cluster $M_{200}$ using two scaling relations of \citet{Munari+13}, the relation of Eq.~(1) in that paper, that is based on the NFW profile, and the relation labelled "AGN gal" in Table~1 of that paper, that is based on the galaxies identified in hydrodynamical simulations with AGN feedback. We obtain $M_{200}=1.3 \pm 0.2 \times 10^{15} \, M_{\odot}$, by considering the average of the values obtained using the two scaling relations. Our $M_{200}$ value is in agreement within 1 $\sigma$ with the virial mass estimate obtained by \citet{Barrena+02}, as well as with the mass estimate obtained from gravitational lensing \citep{CGM04,SF07}.

We run the \texttt{DS+} algorithm in the no-overlapping mode to the data set of 191 member galaxies within the cluster virial radius $r_{200}=2.1$ Mpc. We find ten substructures with a formal probability $p \leq 0.01$. We identify 77 galaxies as members of these substructures, and based on our \texttt{DS+} purity estimate we expect only $\sim 50$ of them to be also members of real groups. It is conceivable that many of the $\sim 27$ spurious members are those assigned to the two substructures of lowest significance, characterized by $p$ values (0.008 and 0.010) much larger than those of the other eight substructures ($\leq 0.003$). 

The properties of the ten detected substructures are listed in Table~\ref{t:Bgroups} and their projected spatial distribution is shown in Fig.~\ref{f:bullet_xyv}. Since we are dealing with small data sets (less than 15 members per substructure) we estimate the substructure velocity dispersions $\sigma_g$ by the gapper method \citep{BFG90,Girardi+93}. Based on our analysis of Sect.~\ref{ss:props}, we expect the group velocity dispersions to be 1.8 times smaller, on average, than the corresponding substructure estimates (see Table~\ref{t:table1}).

Substructure no.~6 in Table~\ref{t:Bgroups} is the Bullet that gives the name to the cluster. Compared to the substructure identified by \citet{Barrena+02}, our substructure mean velocity in the cluster rest frame is slightly below, but still consistent with, that of \citet{Barrena+02}. After correcting the observed value of the substructure velocity dispersion by the average bias factor listed in Table~\ref{t:table1}, we estimate that the Bullet group should be characterized by a velocity dispersion of $472$ km~s$^{-1}$, about twice larger than the estimate of \citet{Barrena+02}.

We apply the above mentioned scaling relations of \citet{Munari+13} to this corrected group velocity dispersion, to estimate a group mass of $1.0_{-0.2}^{+1.1} \times 10^{14} \, M_{\odot}$. This is in agreement with the weak lensing mass estimate by \citet{Bradac+06}, $2.0 \pm 0.2 \times 10^{14}$. The Bullet group to cluster mass ratio we find is $0.07_{-0.06}^{+0.16}$, consistent with the value of 0.1, adopted in the numerical simulation of \citet{SF07}. 

In line with previous analyses \citep{Barrena+02,SF07,MB08} we here consider the classical two-body model to explore the properties of the Bullet collision \citep{GT84,Beers+91}. Taking the mass of the cluster and the group as we inferred from kinematics, the allowed 1 $\sigma$ range of the angle between the collision axis and the plane of the sky is $\sim 4^{\circ}$-$10^{\circ}$ (see Fig.~\ref{f:2body}).
However, the precision of this
estimate is certainly too optimistic, since we have not accounted for the systematic uncertainties inherent to the two-body model.

We then apply the \texttt{MCMAC} code of \citet{Dawson13}. At variance with the classical two-body model, the method developed by \citet{Dawson13} does not assume that the colliding systems are point masses. The cluster and the group are modelled as two spherically symmetric NFW halos. The model assumes energy conservation, zero impact parameter, and that the maximum relative velocities of the two systems is the free-fall velocity given their estimated masses. Dynamical friction is not included in the model. The model is incorporated in a Monte Carlo implementation, wherein parameter values are drawn randomly from observables with associated uncertainties. The observables are the masses of the colliding systems, their mass concentrations, mean redshifts, and the projected distance between the two. 

As before, we adopt the masses and uncertainties we derived from the cluster and group velocity dispersions. We do not measure their mass concentrations, and we, therefore, adopt the mass-concentration relation of \citet{Duffy+08}, that is the internal default of the \texttt{MCMAC} code. We run 50000 Monte Carlo resamplings. In Fig.~\ref{f:mcmac} we show the results as 68\% confidence regions in the plane of "time since the collision" versus "relative 3D velocity at the collision time". The green contour
corresponds to the solution obtained with no external constraint on the angle of the collision. If we discard the solutions with a collision angle outside the range $\sim 4^{\circ}$-$10^{\circ}$ suggested by the two-body model (Fig.~\ref{f:2body}), we obtain the orange contour in Fig.~\ref{f:mcmac}. As explained above, the allowed range for the collision angle that we infer from the two-body model is too restrictive because it ignores systematic uncertainties. Another, possibly more reliable, estimate of the allowed collision angle has been derived by \citet{WCN18}, based on the identification of analogues of observed systems in cosmological n-body simulations. They constrained the collision angle of the Bullet to be $\leq 29^{\circ}$ at the 68\% confidence level. Inserting this constraint in our \texttt{MCMAC} solution gives the red contour in Fig.~\ref{f:mcmac}.

From the \texttt{MCMAC} analysis we conclude that the observational uncertainties are currently too large to allow strong constraints on the geometry, the timing, and the kinematics of the Bullet collision. Our results suggest that the collision occurred within the last 500 Myr, and that the Bullet maximum collision speed was in the range $\sim 2000-4000$ km~s$^{-1}$. The Bullet velocity we find is therefore significantly smaller than the velocity of the bow shock preceding the Bullet \citep[4700 km~s$^{-1}$, ][]{Markevitch06}, as expected from numerical simulations \citep{Milosavljevic+07,SF07}. According to \citet{TDN15}, a collision of two
massive systems such as the Bullet cluster and group, with a 
collision velocity of $\sim 3000$ km~s$^{-1}$ is a rare, but not impossible, event
in a $\Lambda$CDM cosmology. 

Our \texttt{DS+} analysis identifies other (previously unidentified) seven substructures with a \texttt{DS+} probability similar to that of the Bullet (see Table~\ref{t:Bgroups}; we ignore the two substructures of lowest significance in the following discussion). Some have very large velocity dispersion estimates, but are not incompatible with typical group values, given the large error bars and that they are expected to be over-estimated by a bias factor of 1.8, on average (see Table~\ref{t:table1}). 

Substructure no. 3 lies along the Bullet collision axis \citep[as inferred from X-ray images,][]{Markevitch+02}. It has a velocity that is much larger, but compatible within the uncertainties, than that of the Bullet (see Table~\ref{t:Bgroups} and the right-hand panel of Fig.~\ref{f:bullet_xyv}), so it might be originating from the same (as yet unidentified) large scale structure filament whence the Bullet itself came from. The main cluster axis, almost orthogonal to the Bullet collision axis, is traced by four substructures (nos. 2, 7, 4, 1, from bottom left to top right in Fig.~\ref{f:bullet_xyv}). The elongation of the cluster has been suggested to indicate another merger axis for the cluster \citep{LF14,Sikhosana+22}. Our detection of substructures along this axis lends support to this hypothesis, although the lack of a coherent velocity pattern along this (hypothetical) merger axis
(see Fig.~\ref{f:bullet_xyv}, right-hand panel) suggests that multiple episodes of accretion have occurred already along the same axis, with some groups observed before and some after, their pericenter passages.
Substructure no. 5 does not seem to be related to either of the two main collision axes.


\section{Summary and conclusions}
\label{s:conc}
We present a new method for the identification and characterization of group-sized substructures in clusters of galaxies. Our new method, \texttt{DS+}, is based on the positions and velocities of cluster galaxies, and it is an improvement and extension of the traditional method of \citet{DS88}. The method does not provide a global measure of the amount of substructures in a cluster, as most methods do, but it identifies the galaxies that belong to substructures and the substructure themselves. The method can be run in two modes: overlapping and no-overlapping. The former mode allows the most complete identification of galaxies in substructures, while the latter operational mode allows to uniquely identify substructures as independent galaxy associations.

We test \texttt{DS+} on cosmological halos of cluster size extracted from the IllustrisTNG simulation, where infalling groups have been identified by the FoF technique. On average, each of these halos contains 190 galaxies down to a stellar mass of $1.5 \times 10^8 M_{\odot}$.  We find that our method (run in its overlapping mode) successfully identify $\sim 80$\% of the group galaxies as members of substructures, even in groups with less than 10 member galaxies. At least 60\% of the galaxies assigned to the detected substructures are also members of real groups. 

We then compare the properties of the detected substructures 
in the no-overlapping mode of \texttt{DS+}, with those of the matched real groups, by associating to each detected substructure the group with the largest number of common galaxies. We find that the mean velocity, size, velocity dispersion, and stellar mass of the detected substructures, are significantly correlated with the corresponding properties of the matched groups, albeit with a large scatter and a substantial bias. It is then possible to use the properties of the detected substructures to learn about the properties of the real groups, but only on average, by taking into account the biases.

We apply the \texttt{DS+} method to the Bullet cluster, as an example. We find ten significant substructures, one of which corresponds to the group that gives the name to the cluster. We study the geometry and kinematics of the Bullet collision and find consistent results with previous studies, setting 68\% confidence limits to the collision velocity (2000-4000 km~s$^{-1}$) and the collision time ($\lesssim 0.5$ Gyr). The other detected substructures suggest the presence of another collision axis that corresponds to the main South-East to North-West elongation of the cluster. 

We conclude that \texttt{DS+} is a reliable and useful method for the identification of substructure galaxies and substructures themselves in clusters. A Python implementation of our method is freely available for use in GitHub.

\begin{acknowledgements}
      AB thanks Peter Katgert, for his precious collaboration in past years on the development of a method for the detection of cluster substructures. This work was largely supported by the LACEGAL program. JB and MA acknowlegde finantial support from FONCYT, Argentina through PICT 2019-1600.
      This research has made use of the NASA/IPAC Extragalactic Database (NED), which is funded by the National Aeronautics and Space Administration and operated by the California Institute of Technology.
\end{acknowledgements}

\bibliography{ds}


\appendix

\section{Brief description of the use of DS+ public code}
\label{app:public_code}

Here we present a brief description of the \texttt{DS+} code, in the python and public version, available in the GitHub repository\footnote{ \href{https://github.com/josegit88/MilaDS}{https://github.com/josegit88/MilaDS}}. The \texttt{DS+} method has been implemented as the main function into the \texttt{MilaDS} code. Briefly, the principal inputs of the code are the spatial x,y coordinates, in kpc, the line-of-sight velocities, the redshift of the cluster, and, as an option,  the amount of ``re-samplings'' $N_{sims}$ (nsims) that use random samples to assess the probability of the detected substructures, and the upper limit probability (Plim\_P) below which the detections are considered significant.
\\

\texttt{DSp\_groups} is the main function of \texttt{MilaDS}, that receives input information and processes three principal (sequential) stages:
\begin{itemize}
    \item Individual probability of each galaxy to belong to some \texttt{DS+} group of any multiplicity.
    \item allocation of each galaxy only in one \texttt{DS+} group, following the $P_{lim}$ priority. Assign each \texttt{DS+} group one unique group number, so galaxies outside of the final group allocation possess group number $\rm{GrNr}=-1$, and zero in their group properties.
    \item summary of \texttt{DS+} groups properties, such as group number (GrNr), Number of galaxies in each group (Ngal), radial cluster-centric distance (R in kpc), group size (size in kpc), velocity dispersions of the group (sigma km/s), mean velocity of the group (Vmean in km/s), minimum probability of the group (Pmin), an average of individual probabilities of all galaxies in each \texttt{DS+} detected group (Pmin\_avr).
\end{itemize}

The shortest running form of the \texttt{DS+} code, for a cluster located at z=0.296, using 500 re-simulations, and an upper probability limit of 1\%, is as follows:\\





\begin{lstlisting}
# Import MilaDS and other packages:
>>> import milaDS
>>> import numpy as np

>>> my_data = np.genfromtxt("cluster_C1.dat")
... # 0:galaxies IDs
... # 1:X in kpc
... # 2:Y in kpc
... # 3:rest-frame Vel (V_los) in km/s

>>> data_DSp, data_grs_alloc, summary_DSp_grs =
...     milaDS.DSp_groups(
...         Xcoor=my_data[:,1],
...         Ycoor=my_data[:,2],
...         Vlos=my_data[:,3],
...         Zclus=0.296,
...         cluster_name="C1",
...         nsims=500,
...         Plim_P=1 )
\end{lstlisting}

\section{\texttt{DS+} in the accretion history}
\label{app:accretion_hist_app}
Here we present the projected spatial distribution of the galaxies in the same simulated cluster shown in Fig.~\ref{f:distribution}, at different time snapshots, since $z=0$ (top panel) and separated by $\sim 2$ Gyr. Note that the colors are reset at each snapshot, so it would not be entirely correct to track groups along different time snapshot based on their colors.

\begin{figure*}
\centering
\includegraphics[width=\hsize]{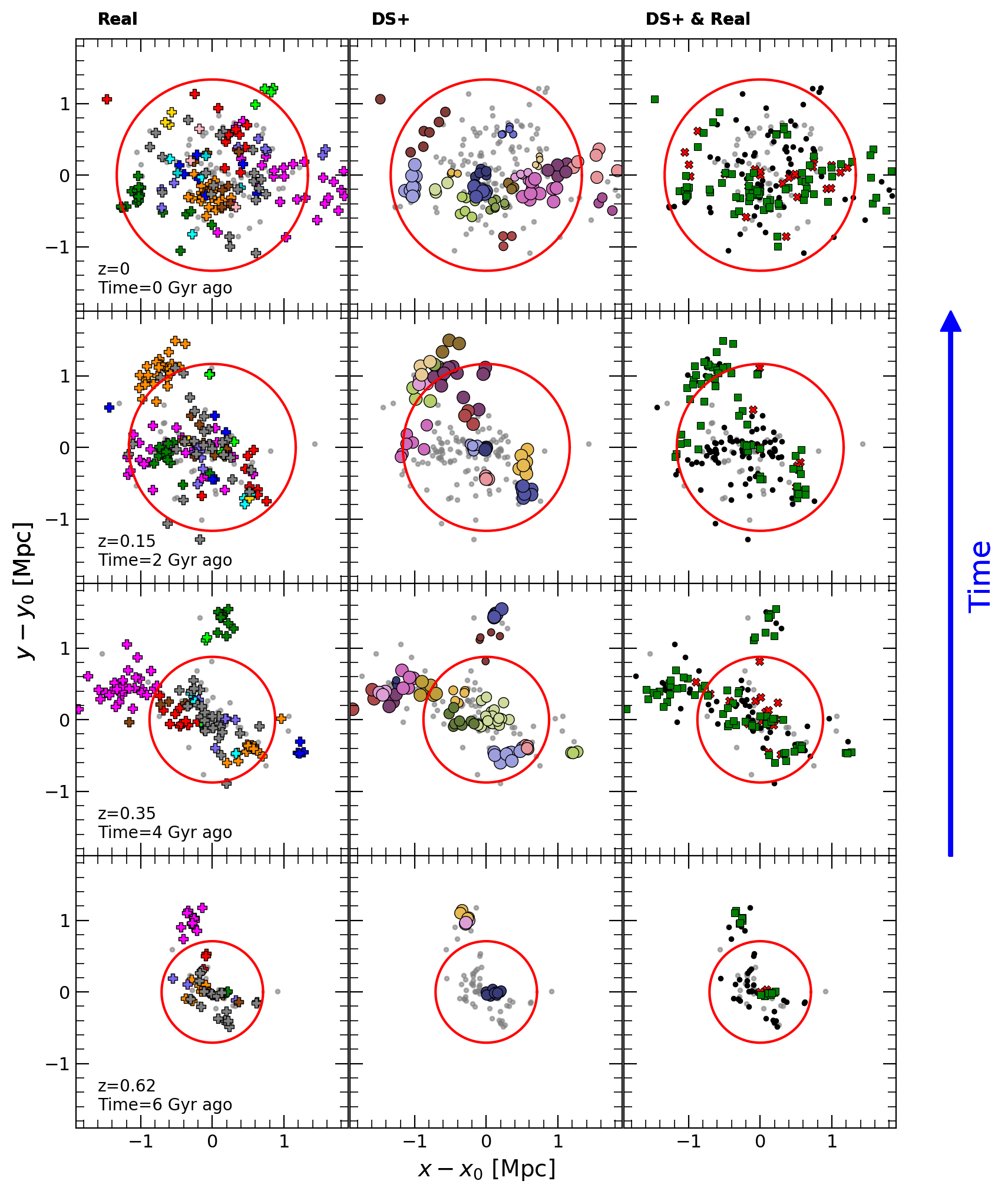}
  \caption{Same cluster presented in
  Fig.~\ref{f:distribution} but for different time snapshots, starting at $z=0$ (top panel) and each $\sim 2$ Gyr look-back in time, until $z \sim 0.62$ (bottom panel).  Coordinates are in Mpc from the cluster center, defined as the position of the particle with the minimum gravitational potential energy, and the red circle indicates the virial radius of the cluster at the corresponding time.} 
     \label{f:distribution_app}
\end{figure*}

\end{document}